\documentclass[twocolumn,10pt]{article}
\usepackage{abstract}

\usepackage{graphicx}

\usepackage[style=ieee,backend=bibtex]{biblatex}
\renewbibmacro*{bbx:savehash}{} % http://tex.stackexchange.com/questions/131625/biblatex-ieee-style-same-authors
\usepackage{inconsolata}
\usepackage{listings}
\usepackage{lstlang0} % Go language support
\begin{document}

\title{An Implementation and Analysis of a Kernel\\Network Stack in Go with the CSP Style}
\author{Harshal Sheth \& Aashish Welling}
\date{}

\twocolumn[
	\maketitle
	\begin{onecolabstract}
Modern operating system kernels are written in lower-level languages such as C.
Although the low-level functionalities of C are often useful within kernels, they also give rise to several classes of bugs.
Kernels written in higher level languages avoid many of these potential problems, at the possible cost of decreased performance.
This research evaluates the advantages and disadvantages of a kernel written in a higher level language. To do this, the network stack subsystem of the kernel was implemented in Go with the Communicating Sequential Processes (CSP) style.
Go is a high-level programming language that supports the CSP style, which recommends splitting large tasks into several smaller ones running in independent ``threads''.
Modules for the major networking protocols, including Ethernet, ARP, IPv4, ICMP, UDP, and TCP, were implemented.
In this study, the implemented Go network stack, called GoNet, was compared to a representative network stack written in C. The GoNet code is more readable and generally performs better than that of its C stack counterparts.
From this, it can be concluded that Go with CSP style is a viable alternative to C for the language of kernel implementations.
	\end{onecolabstract}
]

\section{Acknowledgments}
We would like to thank our mentor, Cody Cutler, for his constant guidance and encouragement. We would also like to acknowledge Prof. Frans Kaashoek, who recommended an initial direction for this project. Finally, we would like to thank the MIT PRIMES program for providing us with this opportunity.

\section{Introduction}
Modern operating systems utilize a kernel to interface with the hardware available to them. Most current operating system kernels are written in the C programming language, which allows them to perform low-level functions effectively. However, C also allows a variety of problems to occur. This paper explores the viability of writing a kernel with CSP style in the Go programming language as a means of avoiding some of the problems associated with current operating system kernels. The network stack, one of many kernel subsystems, was built to evaluate the advantages and disadvantages of this approach. 
To ensure that using Go with CSP style does not hurt the performance of the network stack, the stack's performance was then compared to that of a conventional C language network stack.
The readability, modularity, and concurrency of the two network stacks' code were also evaluated.

\subsection{Operating System Kernels}
Computers are an integral part of modern day society. Computers are expected to be both reliable and efficient. This requires a stable and bug-free operating system kernel, as otherwise, the bugs within the kernel may make other user applications operate unstably and unreliably. 
The operating system kernel serves as a bridge between the applications and users of a computer and the hardware of the machine. The kernel manages the system resources, including memory and hard disk space, and handles the scheduling of processes on the CPUs. It also provides users access to input and output devices and enables network access. User applications run on top of the kernel, and make use of the kernel's functionality through its library of system calls.

\subsection{Problems with Modern Kernels}
Most commodity operating system kernels are implemented in the C programming language. C is the most popular kernel language because it gives a high degree of control over memory usage and other lower level aspects of the program operation. This freedom comes at the cost of allowing problems such as double-free bugs (freeing memory twice), out of bounds errors on arrays (accessing memory that is not part of an array), and deadlocks. It also does not ensure type safety (preventing misinterpretation of data by interpreting data of one type as another type). 

As the number of microprocessor cores per computer increases \cite{Sutter:concurrency}, the ability to take advantage of multithreading is increasingly advantageous to a kernel's overall design.
However, kernels implemented in C are not able to easily take advantage of all of the cores of a machine, because C does not lend itself to leveraging modern microprocessor features. Threads in C, which are used to distribute work among cores, are expensive in both memory and CPU usage; synchronizing these threads is even more difficult and sometimes convoluted.

\subsection{A Kernel Written in a Higher Level Language}

One way to overcome some of these drawbacks is to implement the kernel in a higher level language. This may eliminate many of the problems associated with kernels implemented in C. For example, many higher level languages provide array bounds checking and garbage collection. However, programs that are written in higher level languages generally run slower than those written in C, and sometimes will incur additional overheads from interpreters, automatic memory management, and garbage collection. In addition, the more abstracted higher level languages could make it difficult to perform some of the kernel's low-level tasks.

\subsection{Existing Work}
There have been a few attempts to implement kernels using higher level languages. None of these have achieved widespread adoption for a variety of reasons. 

\paragraph{Mirage}
Mirage is a Linux Foundation project that focuses on turning a web application into a ``standalone, specialized unikernel that runs under the Xen hypervisor'' \cite{MirageOS}. It contains rudimentary implementations of the kernel subsystems, written in OCaml. Because it is built for use within a unikernel, a single-user single-process kernel designed specifically to run in a Virtual Machine, it does not satisfy the needs of most users. In addition, it is not able to achieve parallelism on multiple cores, as it was built for running within a single process.

\paragraph{Pycorn}
Pycorn is an operating system written in Python. It currently is compatible with only 16-bit ARM-based microcomputers \cite{Pycorn}. Because Python is an interpreted language, Pycorn is extremely slow in practice, and performance is not one of the project's goals. Because Pycorn has limited target platforms and is not focused on performance, it has never been fit for widespread use. The project has been inactive since late 2012.

\subsection{Kernel Subsystem}
Since implementing an entire kernel is a massive engineering effort, a single kernel subsystem was implemented instead. 
The subsystem chosen was the network stack, which is a necessary feature of any kernel. The network stack's functionality and performance can easily be tested, making it an ideal subsystem to implement.

\subsection{Programming Language Selection}
A kernel subsystem was built in Go to demonstrate the comparative advantages of writing the kernel in a higher level language. Go, specifically, was chosen because it lends itself to the concurrent sequential processes (CSP) style. The CSP style promotes deconstructing complicated tasks into smaller, more manageable subtasks. These subtasks can be done with individual processes, which communicate with each other to complete the original, larger task. Goals of the CSP style include helping the programmer design, implement, and validate complex computer systems \cite{CSPbook}, and this is especially important when designing software as complicated as a kernel. 
Go provides a thread-safe way of using CSP style through its version of threads, called \textit{goroutines}, and a synchronized communication construct called a \textit{channel}. The Go runtime automatically schedules \textit{goroutines} onto the physical cores of the system.
The CSP style allows the Go programmer to easily take advantage of all of the cores of a computer while maintaining readability and reducing bugs. This is because the network stack can be split into multiple subtasks that can all run inside their own \textit{goroutines}, which are dynamically scheduled to efficiently take advantage of all available cores. This also improves modularity of the code which improves readability and makes it easier to debug.
The CSP style is only feasible in a garbage collected language. Go provides the necessary garbage collection as well as strong typing, which eliminates entire classes of bugs including incorrect type casting, double free errors, and use after free errors. This, among other things, makes Go code simple and easy to maintain. Furthermore, Go's defer statement allows for easier cleanup at the end of functions, reducing the likelihood of deadlocks from neglecting to unlock mutexes. 

\subsection{Purpose}
The advantages of Go and CSP Style may come with a cost. For example, garbage collection has a performance overhead and causes the entire Go runtime to suspend briefly \cite{GolangGCSTW}. In addition, using multiple cores requires communication between these cores, which can be expensive. The purpose of this project is to determine whether the benefits of Go, a higher level language, and CSP style outweigh the disadvantage of having decreased raw speed.

\section{Methods}
To implement a fully independent network stack, the Go stack, named GoNet, was built on the tap interface. For full functionality, all basic network protocols, including Ethernet, ARP, IPv4, ICMP, UDP, and TCP, were implemented. To ensure that GoNet's performance was not impacted, latency and throughput were measured and compared to that of a similar network stack written in C.

\subsection{Tap Interface}
To fully simulate an independent network stack, GoNet operates on a virtual network interface called tap. A tap interface is a virtual network interface, and it mimics actual hardware with simple software. GoNet reads and writes to the tap interface as if it were a normal, physical interface, and the tap interface, in conjunction with the bridge interface, acts as a router into a subnetwork of the host operating system. This allows GoNet to even utilize its own MAC address and IP address, and to connect to external networks.

\subsection{Protocol Implementations}
\begin{figure}
\begin{center}
  \includegraphics[width=0.47\textwidth]{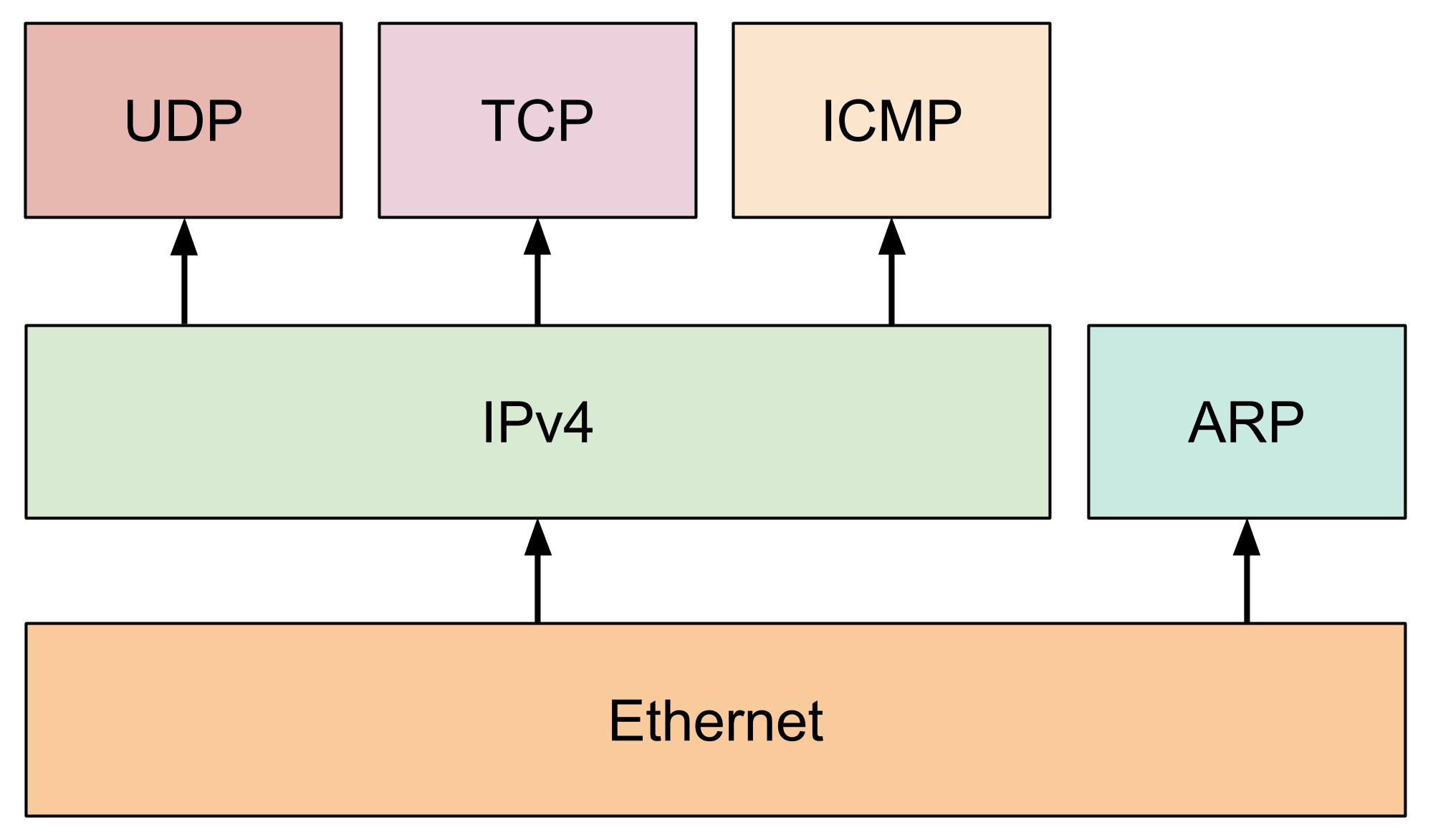}
  \caption{This flowchart illustrates how each network stack protocol receives and processes independently from the other protocols. Each protocol runs in its own set of \textit{goroutines}. Therefore, each protocol can run concurrently with one another. Received data are passed up the stack in \textit{channels} which are represented by the black arrows.}
  \label{fig:parallelization}
\end{center}
\end{figure}

GoNet implements protocols on the data-link, network, and transport layers \cite{OSIRREFRFC}. Each layer runs independently of the other layers and protocols, as shown in Figure \ref{fig:parallelization}. This allows for increased concurrency, as well as increased efficiency under high loads. 

\begin{figure*}
\begin{center}
  \includegraphics[trim=50 30 30 10,clip,width=0.8\textwidth]{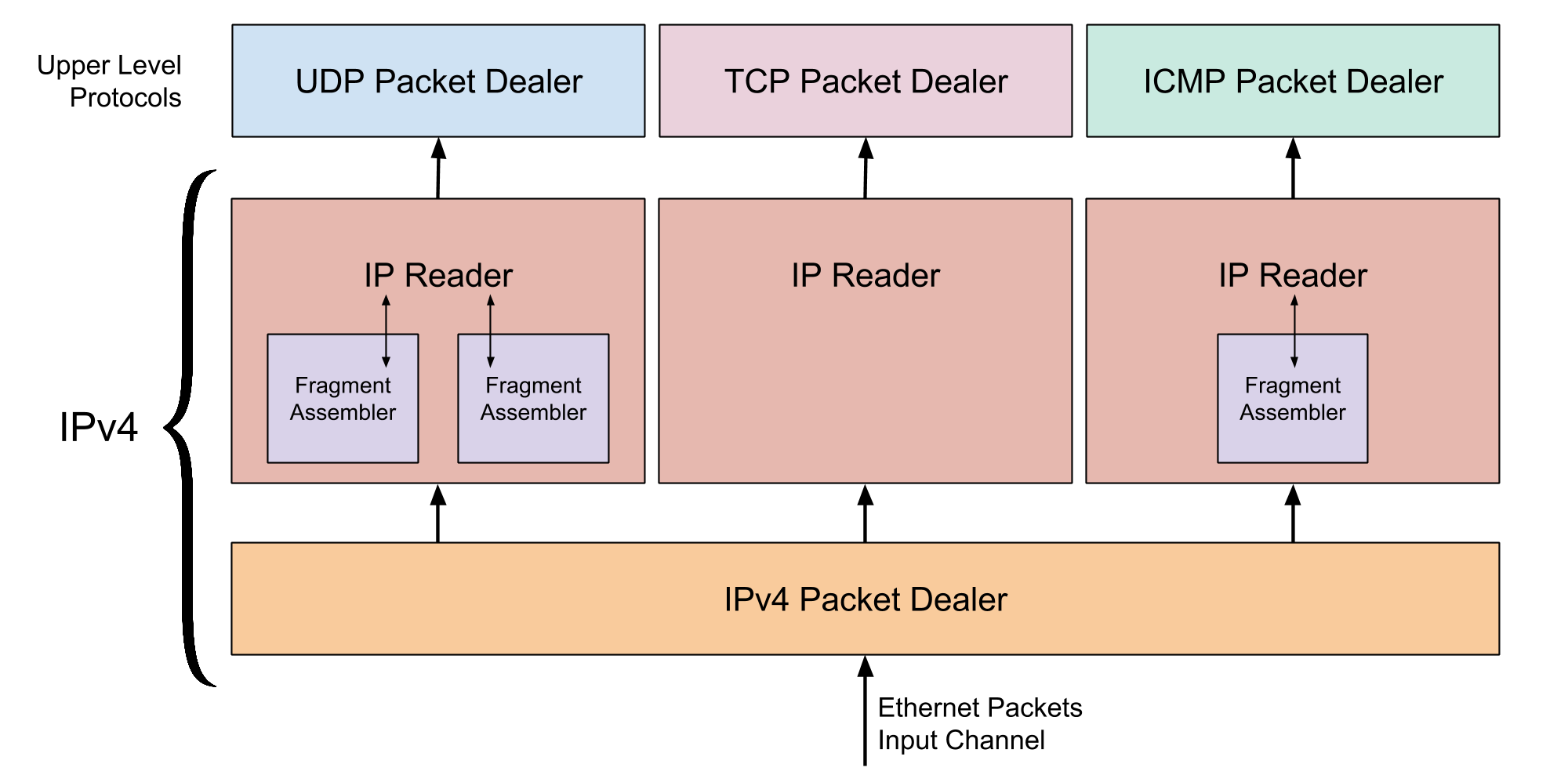}
  \caption{This flowchart shows the design of the IPv4 protocol packet dealer. Each box represents a \textit{goroutine}, and each black arrow represents a \textit{channel}. The IPv4 packet dealer reads packets from the output \textit{channel} of the Ethernet layer and forwards these packets to the correct IP Reader using \textit{channels}. The packets are processed by the IP readers and are then forwarded to the packet dealer of the protocol above.}
  \label{fig:packet-dealer}
\end{center}
\end{figure*}

The implementation of each protocol uses a similar structure: a ``packet dealer''. The IP packet dealer is illustrated in Figure \ref{fig:packet-dealer}. The packet dealer reads packets from the lower layer, transmitted through \textit{channels}. \textit{Channels} are represented by the arrows in Figures \ref{fig:parallelization} and \ref{fig:packet-dealer}. The IP packet dealer sends packets to different IP readers running in their own \textit{goroutines}, represented in figure \ref{fig:packet-dealer} by separate boxes. As IP readers finish processing the packets they receive from the IP packet dealer, they forward the processed data to the next layer packet dealers.

% from https://en.wikibooks.org/wiki/LaTeX/Floats,_Figures_and_Captions#Wide_figures_in_two-column_documents

\paragraph{Ethernet}
The Ethernet layer allows for different network layer protocols to bind to a specific Ethernet protocol. For example, the IPv4 implementation binds to the Ethernet protocol 2048 to receive all IPv4 packets, and the ARP implementation binds to Ethernet protocol 2054.

\paragraph{ARP}
In order to send data to other network stacks on a local network, GoNet needs the MAC address of the target machine. The Address Resolution Protocol (ARP) is implemented to enable GoNet to obtain this information.
ARP allows GoNet to get the MAC address from the destination computer's target protocol address, such as the destination computer's IP address \cite{ARPRFC}. The GoNet implementation of ARP creates a \textit{goroutine} for each ARP request. This allows each ARP request \textit{goroutine} to block until either the main ARP packet dealer notifies it of a response or the request times out. 

\paragraph{IPv4}
The Internet Protocol Version 4 (IPv4) design is illustrated in Figure \ref{fig:packet-dealer}. As explained earlier, it uses a packet dealer structure. It also includes multiple IP Readers, and fragment assemblers when needed. Communication between all of these components is accomplished through the \textit{channels} represented by the black arrows in the figure. 

IPv4 fragmentation is utilized when an IP payload would make the IP packet larger than the maximum transmission unit (MTU). The IP segment is split up into multiple fragments, each containing data needed for reassembly. When the segments at the destination, they must be reassembled into the original IP segment \cite{IPRFC}.
GoNet's fragment assemblers illustrate the advantages of CSP style.

Each fragment assembler encapsulates both the process of reassembling a fragmented IP segment and the associated data. This approach reduces the complexity of the code, because each fragmented IP packet has its own designated assembler, in the style of CSP. This is contrary to traditional fragment assembly methods, where a global data structure manages the fragment assembly for all packets. This localization of data is made more feasible by the lightweight design of \textit{goroutines} and the garbage collected Go language, which greatly reduces the likelihood of memory leaks.

\paragraph{ICMP and Ping}
GoNet implements the ping portion of the Internet Control Message Protocol (ICMP). The ICMP implementation follows the normal packet dealer structure. The ping implementation also has its own packet dealer which handles all of the ICMP ping packets sent first to the ICMP packet dealer. The ping packet dealer forwards received ping requests to a special set of \textit{goroutines} that reply to ping requests. If GoNet has sent ping requests, then the ping packet dealer forwards the replies to a dedicated \textit{goroutine} that is started for each of the mentioned ping requests.

\paragraph{UDP}
The User Datagram Protocol (UDP) is a connectionless protocol. Because UDP is a relatively simple protocol, the GoNet implementation just uses a basic packet dealer to forward packets to their associated UDP readers.

\paragraph{TCP}
The Transmission Control Protocol (TCP) is a connection-oriented transport layer protocol that guarantees in-order delivery of data. Because TCP is connection-oriented, it utilizes a server and a client to initialize a connection. Once a connection is established, it is managed by a Transmission Control Block (TCB) \cite{TCPRFC}. 

The GoNet implementation of TCP uses the standard packet dealer structures to manage source and destination ports. 
Each TCB utilizes two long-running \textit{goroutines}. One processes incoming packets. The other waits for and sends data, as well as creates additional \textit{goroutines} that manage the retransmission of single packets. Each of these two long-running \textit{goroutines} represents half of the duplex TCP connection. Internally, it also uses \textit{channels} to synchronize and manage all of the \textit{goroutines} that are created. For example, the incoming packet processor \textit{goroutine} uses \textit{channels} to notify packet retransmission \textit{goroutines} when an acknowledgment packet arrives.

\subsection{Testing}
GoNet's performance was compared to that of tapip, a multi-threaded network stack written in C \cite{tapip}. This comparison allows for the evaluation of the pros and cons of a network stack written in a higher level language with the CSP style. Both stacks implement similar protocols, operate in user space, and utilize a tap interface. This allows the performance of both stacks to be compared fairly. The testing was performed on a Ubuntu 14.04 machine with Linux 3.13.0, 16 GB of memory, and an Intel Xeon Quad-Core Dual Socket processor. 

\subsubsection{Latency}
The first performance metric that was evaluated was latency. To measure latency, the response times of 50 ping requests were averaged. The ping requests were sent from the Linux kernel that both stacks were running on.
To determine the stacks' performance under increased load, multiple pings were sent from the Linux kernel simultaneously. The test was run from 1 to 1000 concurrent ping ``connections'' to simulate possible loads that a network stack might endure.
To ensure that the tests on the two stacks were run fairly, all other variables were held constant, including the number of ping requests each ping ``connection'' would send, the ICMP receive buffer size, the interval between the ping requests, and the ping request packet size.

\subsubsection{Throughput}
The second performance metric that was evaluated was throughput. The throughput of a stack is the amount of data that it can send or receive in a given amount of time.
The following process was used to measure the throughput of the stacks:
\begin{enumerate}
\item A TCP server was initialized.
\item A TCP client was initialized. The connection was made over the local network (localhost) to eliminate any overhead caused by the tap interface. 
\item Four kilobytes (kB) of data were sent from the client to the server. 
\item The total real time that the stack took to complete the said procedures completely was measured. This time, along with the specific amount of data sent, was used to calculate throughput.
\end{enumerate}
The stacks' performances were measured as the number of clients increased, to test the comparative scalability of the stacks. The test was done up to 100 concurrent clients.

A variety of precautions were taken to ensure that the throughput was measured precisely. For example, all comparable buffer sizes were set equal. In tapip, each client and server connection ran in its own thread; GoNet was similar, except it used \textit{goroutines} instead of threads. Additional precautions were taken to ensure that each connection had completed before stack termination and that the payloads of each connection were transferred in their entirety. 

\section{Results and Discussion}
The code of GoNet was far simpler than that of its C stack counterpart. In addition, the performance, over both latency and throughput, of GoNet was actually better than that of tapip. 

\subsection{Correctness}
In terms of protocol operation, both GoNet and tapip were correct. This was determined by successfully testing both stacks against a Linux Kernel TCP endpoint. However, tapip leaked memory during the test. This is because tapip stores packets in packet buffers and these buffers are sometimes double freed or not freed at all. When tapip double frees memory, it either crashes or causes undefined behavior. When tapip does not free memory, the unfreed memory accumulates and hogs resources until the system eventually crashes. Go makes it easy to avoid these types of problems with its built-in garbage collection.

\subsection{Code Comparison}
It is hard to quantitatively evaluate the merits of writing code in the Go language compared to the C language. The following code comparisons are used to illustrate some of the advantages of higher level languages. The code that is being compared is all part of the IP fragment reassembly process. The C code is on the top of each comparison segment, and the Go code is on the bottom.

\paragraph{Fragment Reassembly Initialization}
The following code segments compare the steps that tapip and GoNet take to initialize a new fragment reassembler when a new fragmented IP segment arrives. GoNet creates a new \textit{goroutine} for each packet that is being reassembled while tapip uses a global structure to hold the data for all of the ongoing reassemblies. 

\begin{minipage}{0.47\linewidth}
\begin{lstlisting}[language=C]
struct fragment *frag;
frag = xmalloc(sizeof(*frag));
list_add(&frag->frag_list, &frag_head);
list_init(&frag->frag_pkb);
return frag;
\end{lstlisting}
\end{minipage}
\hrule
\begin{minipage}[t]{0.47\linewidth}
\begin{lstlisting}[language=Go]
ipr.fragBuf[bufID] = make(chan []byte, 
			FRAG_ASSEM_BUF_SZ)

quit := make(chan bool, 1)
done := make(chan bool, 1)
didQuit := make(chan bool, 1)

go ipr.fragAssembler( /* ... */ )
go ipr.killFragAssembler( /* ... */ )
\end{lstlisting}
\end{minipage}
\vspace{-4mm}

\paragraph{Adding Fragments to a Reassembly Queue}
These code comparisons show how the structure defined in the fragment reassembly initialization makes adding fragments to the processing queue easier in GoNet than in the C stack. This allows the \textit{goroutine} that processes an IP segment in GoNet to simply forward the packet to its respective reassembler and move onto subsequent packets. This improves the modularity of GoNet's code, as well as its readability and concurrency. 

\begin{minipage}{0.47\linewidth}
\begin{lstlisting}[language=C]
int insert_frag(/* ... */) {
  /* additional fragment processing */
  list_add(&pkb->pk_list, pos);
  return 0;
frag_drop:
  free_pkb(pkb);
  return -1;
}
\end{lstlisting}
\end{minipage}
\hrule
\begin{minipage}[t]{0.47\linewidth}
\begin{lstlisting}[language=Go]
ipr.fragBuf[bufID] <- b
\end{lstlisting}
\end{minipage}
\vspace{-4mm}

\paragraph{Dealing With Completed Fragments}
The following code segments underscore the advantages that CSP style and the Go language provide over current C stacks.
Tapip has to deal with fragmented packets before it can move on to subsequent packets. This introduces a variety of problems. For example, it forces the C IP implementation to track the states of all of the ongoing fragment reassemblies at the same time. This encourages the use of possibly complicated global variables and structures and makes thread synchronization difficult. 
In contrast, GoNet spawns a separate fragment assembler \textit{goroutine} for each new fragmented IP packet that it receives. Each \textit{goroutine} is responsible for all of the separate fragments that make up the IP segment. After the fragment assembler finishes assembling the packet, it simply sends the reassembled segment back into processing. This process is completely independent of the main IP packet processing \textit{goroutines}, and hence allows for concurrency and parallelism, as well as far more readable, understandable, and clean code. 

\begin{minipage}{0.47\linewidth}
\begin{lstlisting}[language=C]
if (complete_frag(frag))
  pkb = reass_frag(frag);
else pkb = NULL;
return pkb;

struct pkbuf *reass_frag(
                struct fragment *frag) {
  /* more processing */
  delete_frag(frag);
  return pkb;
}
\end{lstlisting}
\end{minipage}
\hrule
\begin{minipage}[t]{0.47\linewidth}
\begin{lstlisting}[language=Go]
ipr.incomingPackets <- append(
                fullPacketHdr, payload...)
done <- true
\end{lstlisting}
\end{minipage}
\vspace{-4mm}

\paragraph{Fragmentation Cleanup}
Both stacks have to delete an entry from a data structure. The data structure tracks \textit{channels} for Go and defragmentation structures for tapip. In addition, tapip has to explicitly free the memory allocated for each fragmented packet buffer, as well as the memory from the defragmentation structure as well. 

\begin{minipage}{0.47\linewidth}
\begin{lstlisting}[language=C]
struct pkbuf *pkb;
list_del(&frag->frag_list);
while (!list_empty(&frag->frag_pkb)) {
  pkb = frag_head_pkb(frag);
  list_del(&pkb->pk_list);
  free_pkb(pkb);
}
free(frag);
\end{lstlisting}
\end{minipage}
\hrule
\begin{minipage}[t]{0.47\linewidth}
\begin{lstlisting}[language=Go]
delete(ipr.fragBuf, bufID)
\end{lstlisting}
\end{minipage}
\vspace{-4mm}

\subsection{Performance Comparison}
The latency and throughput of both the C stack and GoNet were measured and compared.

\subsubsection{Latency}
\begin{figure}[hb]
\begin{center}
\includegraphics[trim=30 15 100 15,clip,width=0.47\textwidth]{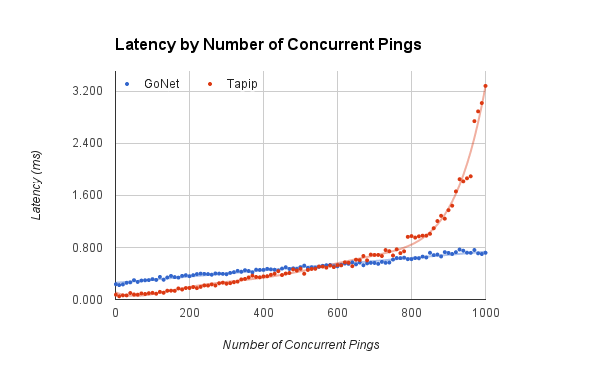}
  \caption{This graph displays the latency of both GoNet and tapip by the number of concurrent pings.}
  \label{fig:latency}
\end{center}
\end{figure}

The trends of the latency test results can be seen in Figure \ref{fig:latency}. The drop rates of both stacks were negligible. With 1 ping, tapip outperformed GoNet by over three times with a latency of 0.074~ms when compared to GoNet's latency of 0.234~ms. However, with 1000 concurrent pings, GoNet outperformed tapip by almost five times with a latency of 0.717~ms when compared to tapip's latency of 3.279~ms. GoNet begins to outperform tapip when the number of concurrent connections becomes greater than about 600. GoNet's latency grows linearly while tapip's latency to grow exponentially. GoNet's latency trend is superior to tapip's latency trend, because, at low numbers of concurrent pings, the latencies of both stacks are small enough to be negligible, but at higher numbers of concurrent pings, the absolute difference in latency is much larger.

Based on these results, it can be inferred that tapip can process a small number of packets very fast while it is slower with processing larger numbers of packets. This is likely because it is not as concurrent as GoNet. In contrast, GoNet takes a longer time to process each packet, but is mostly unaffected by increased load, likely because of the degree of concurrency within the implementations of each protocol. This can be seen in Figure \ref{fig:latency}, as tapip's latency grows much faster than GoNet's, even though it begins with much lower latency.

The sharp increase in the latency of tapip also supports this explanation of the results. After about 800 concurrent connections, tapip becomes unable to field each set of pings requests before the next set is sent by the concurrent ping connections, and hence, a backlog of ping requests develops. This causes a delay in the response to all of the pings, resulting in a sharp growth in tapip's latency. Since tapip does not drop any packets, it is not possible that it is dropping packets because of a full buffer. 

This sharp increase in tapip's response times highlights the underlying problem with its architecture, and the architecture of many other networks stacks as well: processing a single packet at all layers before moving onto a new packet is suboptimal, as it can not scale or achieve parallelism effectively. 

\subsubsection{Throughput}
\begin{figure}
\begin{center}
  \includegraphics[trim=35 15 100 15,clip,width=0.47\textwidth]{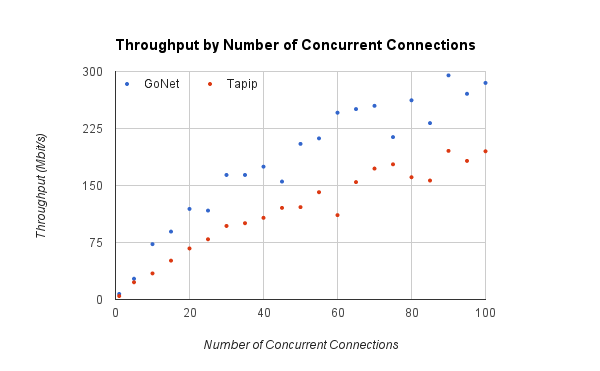}
  \caption{This graph displays the throughput of both GoNet and tapip by the number of concurrent TCP connections. }
  \label{fig:throughput}
\end{center}
\end{figure}

The results of the throughput test can be seen in Figure \ref{fig:throughput}. With 1 concurrent connection, GoNet outperformed tapip with a throughput of 7.3~Mbit/s when compared to tapip's throughput of 4.6~Mbit/s. With 100 connections, GoNet outperformed tapip with a throughput of 284.9~Mbit/s when compared to tapip's throughput of 195.0~Mbit/s. In addition, GoNet's throughput increases at a faster rate than tapip's throughput. This shows that GoNet can continue to scale for even larger numbers of connections while tapip may not be able to handle such load. 

These results make sense given the architecture of GoNet. All of the TCBs in tapip are managed by a single thread. In contrast, each TCB in GoNet has two threads managing it: one for each half of the duplex connection. In this way, GoNet is able to efficiently multiplex a large number of connections onto a limited number of cores more efficiently than tapip. Hence, it achieves far greater throughput for large numbers of connections. With small numbers of connections, GoNet is still slightly more efficient, as GoNet splits the work of the TCB among two \textit{goroutines}, while tapip has one thread performing processing. GoNet outperforms tapip for all numbers of concurrent connections. 

\section{Conclusions}
The operating system kernel is important for managing a computer system's resources. Therefore, the kernel needs to be well designed in order to support the rest of the operating system properly. Modern kernels are written in lower-level languages such as C, which allow several classes of bugs to occur; writing a kernel in a higher level language can eliminate several of these.
However, higher level languages have their own downsides. The network stack, a kernel subsystem, was built in Go to demonstrate the advantages of a kernel written in a higher-level language.
% methods
GoNet and tapip both operate on the tap interface. Both network stacks also implement similar protocols such as Ethernet, IPv4, ARP, UDP, and TCP. The network stack that was built in Go, called GoNet, performs competitively against a similar network stack written in C, called tapip. 

% results
GoNet's code was simpler than that of the C stack, as demonstrated in the IP fragment reassembler code comparison. GoNet, which was built with the CSP style, could simplify and modularize in a more effective way than the C stack. This also allowed for increased concurrency and parallelism and helped improve the performance of GoNet.
In latency tests, GoNet achieved lower latency than tapip for numbers of connections greater than about 600. In throughput testing, its parallelism allowed it to outperform tapip for all numbers of concurrent TCP connections ranging from 1 to 100.

However, there are possible sources of error in the tests. For example, tapip is not a mainstream network stack, and may have room for optimization. Also, the Linux kernel might have scheduled each run of the test differently, which would lead to variation in the results. In addition, the latency test results for both GoNet and tapip could have been limited by the speed of the tap interface that the packets were sent and received on. There also could be other external variables that are unaccounted for that could affect the results of either test. 

There are also alternate explanations for the results of the tests. For latency, unforeseen uncontrolled variables may have caused tapip to slow for larger numbers of connections. 
In addition, tapip leaks memory by not freeing the memory before deleting references to it. For high numbers of concurrent ping connections in the latency test, the higher memory usage of tapip could increase the overhead of tracking allocated blocks of memory and slow the overall program. 
These possibilities are unlikely as they would have caused a more gradual deterioration in tapip's performance rather than the more sudden drop.

This experimentation shows that a kernel subsystem written in Go with CSP style can improve readability, modularity, concurrency, reliability, and stability without significantly affecting performance adversely. This shows that the Go language with CSP style is a viable alternative to the C language for kernel implementations. 

\subsection{Future Work}
This project can be expanded in many different directions. Possible directions:
\begin{itemize}
\item Support could be added for IPv6 in both the transport layer protocols and the network layer protocols \cite{OSIRREFRFC}. This would simply make GoNet more applicable in different environments. 
\item A socket API could be built on top of the existing stack, as this would allow application layer protocols to be built on top of GoNet, extend the functionality of the current stack's implementation, and make the stack POSIX compliant. 
\item The application layer protocols, which are protocols that run on top of UDP, TCP, and other transport layer protocols, could be implemented. Some possible protocols include Secure Shell (SSH), Telnet, the Hypertext Transfer Protocol (HTTP), the File Transfer Protocol (FTP), the Domain Name Service (DNS), and the Network Time Protocol (NTP). Implementing these protocols would allow GoNet to become more functional to end users, and hence could become more ready for use as a user-space alternative to the system network stack. 
\item More detailed CPU and memory profiling could be done to find and remove any bottlenecks in GoNet. Also, race detection and memory analyzers could be used to find any additional problems in GoNet.

\item Additional performance metrics could be developed in order to better understand the differences of the two stacks.

\item Implement other kernel subsystems with the eventual goal of implementing the entire kernel in Go. Moving GoNet into kernel space would also allow for testing when compared to a wider variety of network stacks, as it would become comparable to a wider variety of kernels. In addition, the other kernel subsystems could be compared in the subsystem's proper metric, giving a more holistic view of the advantages and disadvantages of Go with CSP Style. 
\end{itemize}

\printbibliography

\end{document}